\PassOptionsToPackage{table,xcdraw}{xcolor}
\documentclass[11pt,a4paper]{article}
\pdfoutput=1
\usepackage{jheppub}
\usepackage{gensymb}
\DeclareSymbolFont{matha}{OML}{txmi}{m}{it}% txfonts
\DeclareMathSymbol{\varv}{\mathord}{matha}{118}
\usepackage{amsmath}
\usepackage{amssymb}
\usepackage{graphicx}
\usepackage{subfigure}
\usepackage{pstricks}
\usepackage{bm}
\usepackage{pbox}
\usepackage{placeins}
\usepackage{booktabs}
\usepackage[T1]{fontenc}
\usepackage{footnote}
\usepackage{pdfpages}
\usepackage{hhline}
\usepackage{multirow}
\usepackage{multicol}
\usepackage{enumitem}
\usepackage[toc,page]{appendix}
\allowdisplaybreaks
\usepackage{comment}
\usepackage{float}
\usepackage{slashed}
\usepackage{xcolor}
\usepackage{textcomp}
\usepackage{gensymb}
\usepackage{multirow}
\usepackage[numbers]{natbib}
\usepackage{notoccite}

\usepackage{rotating}
\usepackage{tabularx}
\usepackage[labelfont=bf]{caption} % optional

\FloatBarrier

\usepackage[many]{tcolorbox}

\renewcommand{\thetable}{\Roman{table}}
\newcommand\scalemath[2]{\scalebox{#1}{\mbox{\ensuremath{\displaystyle #2}}}}

\newcommand{\e}{\textcolor{red}{\epsilon_1}}
\newcommand{\ee}{\textcolor{blue}{\epsilon_2}}
\newcommand{\es}{\textcolor{red}{\epsilon_1^2}}
\newcommand{\ees}{\textcolor{blue}{\epsilon_2^2}}

\definecolor{MyDarkBlue}{rgb}{0.1, 0.1, 0.8} 
\definecolor{MyLightBlue}{rgb}{0.22,0.51,0.9}
\definecolor{MyGreen}{rgb}{0.0, 0.5, 0.0}
\definecolor{BrickRed}{rgb}{0.8, 0.25, 0.33}

\RequirePackage{hyperref}
\hypersetup{colorlinks, citecolor=MyGreen,linkcolor=cyan, urlcolor=MyLightBlue}

\makeatletter
\gdef\@fpheader{}
\makeatother
\begin{document}
%\vspace*{-0.2in}
\preprint{OSU-HEP-20-10}
\title{\bf 
Flavor Hierarchies from Clockwork in $\boldmath{SO(10)}$ GUT}
\author[]{K.S. Babu,}
\author[]{Shaikh Saad}

\affiliation[]{Department of Physics, Oklahoma State University, Stillwater, OK 74078, USA}

\emailAdd{babu@okstate.edu, shaikh.saad@okstate.edu}
%%%%%%%%%%%%%%%%%%%%%%%%%%%%%
\abstract{
The clockwork mechanism, which can naturally explain the origin of small numbers, is implemented in $SO(10)$ grand unified theories to address the origin of hierarchies in  fermion masses and mixings.  We show that a minimal Yukawa sector involving a $10_H$ and $\overline{126}_H$ of Higgs bosons, extended with two clockwork chains consisting of $16+\overline{16}$ vector-like fermions, can explain the hierarchical patterns with all the Yukawa couplings being of order one. Emergence of a realistic mass spectrum does not require any symmetry that distinguishes the three generations. We develop clockwork-extended $SO(10)$ GUTs both in the context of SUSY and non-SUSY frameworks. Implementation of the mechanism in non-SUSY scenario assumes a Peccei-Quinn symmetry realized at an intermediate scale, with the clockwork sector carrying non-trivial charges, which solves the strong CP problem and provides axion as a dark matter candidate.    
}


%\arxivnumber{}

\maketitle

%%%%%%%%%%%%%%%%%%%%%%%%%%%%%%%%%%%%%%%%%%%%%%%
%%%%%%%%%%%%%%%%%%%%%%%%%%%%%%%%%%%%%%%%%%%%%%%
\section{Introduction}\label{sec:Intro}
The origin of the observed hierarchies in the masses and mixings of quarks and leptons  is a longstanding puzzle in the Standard Model (SM) of particle physics, commonly referred to as the \textit{flavor puzzle}.  Whereas the charged fermion masses and mixings appear to be strongly hierarchical between generations, in the neutrino sector only a mild hierarchy is realized. Additionally, a stronger hierarchy is observed in the up-quark mass ratios compared to the down-quark and charged lepton counterparts, which have very similar patterns.  The SM fails to provide an explanation of the flavor puzzle
as it simply accommodates the observed masses and mixings in terms of completely free Yukawa coupling parameters.  Besides, neutrinos are strictly massless in the SM, which contradicts observations.  There have been many attempts to address the  flavor puzzle among which grand unified theories (GUTs) \cite{Pati:1974yy, Georgi:1974sy, Georgi:1974yf, Georgi:1974my, Fritzsch:1974nn} based on $SO(10)$ gauge group   \cite{Georgi:1974my, Fritzsch:1974nn} are very attractive candidates. In addition to unifying the strong, weak and electromagnetic forces into a single force, in $SO(10)$ GUTs, quarks and leptons of each family are unified into a single irreducible $16$-dimensional  representation. Along with the SM fermions, this $16$-dimensional spinor representation contains the right-handed neutrino that naturally leads to small but non-zero neutrino masses via the seesaw mechanism \cite{Minkowski:1977sc,Yanagida:1979as,Glashow:1979nm,GellMann:1980vs,Mohapatra:1979ia}. The unification of all fermions of a family into an irreducible representation is a good starting point to address the flavor puzzle, owing to the various correlation it provides.   Further appealing features of this theory include a natural understanding of electric charge quantization, automatic anomaly cancellation, and gauge coupling unification at high energy scale around $10^{16}$ GeV  with or without  supersymmetry (SUSY).

Because of  the unification of quarks and leptons of each family into a single irreducible representation, $SO(10)$ GUT is one of the best frameworks that can shed some light on the flavor puzzle.  The Higgs fields that can generate fermion masses at the renormalizable level can be identified from the fermion bilinears
\begin{equation}
16 \times 16 = 10_s + 120_a + 126_s\;, \label{bilin}
\end{equation}
where the subscripts $s$ and $a$ denote symmetric and antisymmetric flavor structures. A minimal Yukawa sector in SUSY 
$SO(10)$ GUT and/or non-SUSY $SO(10)$ with a $U(1)$ Peccei-Quinn (PQ) symmetry \cite{Peccei:1977hh} consists of two Higgs multiplets $10_H$ and $\overline{126}_H$ having the following interactions  \cite{Babu:1992ia}:
\begin{align}
{\cal L}_Y=16^T\left(Y_{10}\,10_H+Y_{126}\overline{126}_H\,\right)16\;.    \label{yuk}
\end{align}
In the non-SUSY framework, the PQ symmetry would require complexification of the $10_H$.  The complex conjugate of $10_H$ however,  does not couple to fermions, owing to its PQ charge, thus preserving the form of Eq. (\ref{yuk}).  The flavor phenomenology would thus be similar in the SUSY version as well the PQ-symmetric non-SUSY version, with the Yukawa couplings given by Eq. (\ref{yuk}) in both cases. Introduction of the PQ symmetry is of course highly motivated on independent grounds, since it naturally solves the strong CP problem and also provides a viable dark matter candidate in the axion.

The fermion mass matrices derived from the minimal Yukawa sector of Eq. (\ref{yuk}) has only 19 real parameters (of which 7 are phases) to fit 19 observed quantities, making the theory very predictive in the flavor sector. These observed quantities are the 6 quark and 3 charged lepton masses; 2 neutrino mass-squared differences; 3 quark mixing angles; 3 lepton mixing angles; and 2 Dirac CP phases -- 1 in the quark mixing matrix and 1 in the lepton mixing matrix. Since these parameters are all interrelated in $SO(10)$ and since they should fit different hierarchical patterns in the charged fermion and the neutral fermion sectors, finding an acceptable fit is a highly non-trivial task. This issue has been been extensively studied  in the literature and the consistency of the minimal Yukawa sector established  \cite{Babu:1992ia,Bajc:2001fe,Bajc:2002iw,Fukuyama:2002ch,Goh:2003sy,
Goh:2003hf,Bertolini:2004eq, Bertolini:2005qb, Babu:2005ia,Bertolini:2006pe, Bajc:2008dc,
Joshipura:2011nn,Altarelli:2013aqa,Dueck:2013gca, Fukuyama:2015kra, Babu:2018tfi, Babu:2018qca, Ohlsson:2019sja}.  In particular, the reactor neutrino mixing angle $\theta_{13}$ was predicted to be large and  close to its experimentally measured value.  While the minimal Yukawa sector involving $10_H + 126_H$ of Higgs fields is sufficient to explain
observed data, a flavor-antisymmetric $120_H$ Yukawa coupling matrix could be added to Eq.\ \eqref{yuk}
\cite{Dutta:2004hp, Dutta:2005ni, Lavoura:2006dv, Ferreira:2015jpa, Fukuyama:2016vgi, Babu:2016bmy, Deppisch:2018flu}. Although the clockwork mechanism we develop here can be straightforwardly extended to this non-minimal case, we do not purse it here.

While achieving good fits to the fermion masses and mixings with the minimal Yukawa sector is certainly a great success for $SO(10)$ GUT, this framework does not explain the hierarchical features observed in the spectrum. One idea that has been pursued to explain the hierarchy is to use additional symmetries that distinguish families which can lead to highly regulated fermion mass matrices (for reviews see Refs. \cite{Babu:2009fd,Feruglio:2015jfa, Xing:2019vks} and  references therein).
In such attempts, typically, the vacuum expectation values (VEVs) of certain flavon fields which break the flavor symmetry have to be arranged in a preferred pattern, which may not be entirely satisfactory.  
Another widely used approach is the implementation of the  Froggatt-Nielsen mechanism \cite{Froggatt:1978nt} with a flavor-dependent $U(1)$ symmetry  wherein lighter fermion masses arise as higher dimensional operators, which are thereby naturally suppressed. While the coefficients of the higher dimensional operators can all be of order one, being non-renormalizable in nature, this setup does not provide quantitative predictions.

Recently, an interesting mechanism dubbed as the  \textit{clockwork mechanism} has been proposed \cite{Choi:2015fiu,Kaplan:2015fuy}  to explain small numbers.  While the initial motivation was to explain the gauge hierarchy problem in the context of relaxions,  this mechanism has been generalized for broader model building purposes in Ref. \cite{Giudice:2016yja}. The clockwork mechanism is an economical and elegant way to naturally generate large hierarchies between different quantities within a theory that contains only $\mathcal{O}(1)$ couplings. Briefly, this mechanism when applied to the flavor puzzle works as follows. The SM is enlarged with $N$ fermions that are vector-like under the SM which may be viewed as a one dimensional lattice.  Due to an enforced symmetry, only nearest-neighbor interactions are allowed among the lattice sites, and only along one direction. Upon integrating out these heavy states on the lattice, hierarchy factors are generated -- which may be even exponential for large number of lattice sites --  via a sharp localization of the zero mode (the SM fermion) towards the boundary of the  lattice. This idea can be readily extended  to $3\times 3$ matrices in flavor space to create inter-generational hierarchies. For implementation of the clockwork mechanism to address flavor puzzle in various contexts see Refs.  \cite{vonGersdorff:2017iym, Patel:2017pct, Alonso:2018bcg, Sannino:2019sch, Smolkovic:2019jow, deSouza:2019wji, vonGersdorff:2020ods}, and for attempts to generate small neutrino masses and mixings  see Refs.\  \cite{Ibarra:2017tju, Banerjee:2018grm, Hong:2019bki, Kitabayashi:2019qvi, Kitabayashi:2020cpo}.  In Refs.\ \cite{vonGersdorff:2017iym, Patel:2017pct, Alonso:2018bcg, Smolkovic:2019jow, deSouza:2019wji}, clockwork mechanism was incorporated into the SM flavour sector by introducing a set of vector-like fermions, where the number of added vector-like particles for each sector (up-type quark, down-type quark, and charged lepton) controls the pattern of the hierarchy for that particular sector. Ref.\ \cite{Alonso:2018bcg} focused on TeV scale clockwork states, where a detailed phenomenology of vector-like fermions is also presented. Moreover, in Ref.\ \cite{Smolkovic:2019jow}, two anomaly free versions of clockwork models with $U(1)$ and $[U(1)]^3$ flavor symmetries  are presented to explain the observed hierarchies in the flavor sector.  By allowing Yukawa couplings of the vector-like states with the Higgs fields that break the GUT symmetry,  in Ref.\ \cite{vonGersdorff:2020ods} flavor hierarchies are addressed in the context of $SU(5)$ and $SO(10)$ GUTs in the SUSY framework. Furthermore, by adding a set of SM singlet fermion states, flavor structures in the neutrino spectrum are analyzed in Ref.\ \cite{Ibarra:2017tju, Hong:2019bki, Kitabayashi:2019qvi}. The observed  flavor structure for the neutrino mass matrix with scalar clockwork has been implemented in Refs.\  \cite{Banerjee:2018grm, Kitabayashi:2020cpo}.

In this work we present an interesting  explanation to the flavor puzzle by combining the minimal Yukawa sector of $SO(10)$ GUT, Eq. (\ref{yuk}), with clockwork  mechanism.  Within our framework, the GUT symmetry relates different flavors, whereas clockwork chains assist in providing  the required hierarchical patterns. Thus the predictivity of the minimal Yukawa sector is preserved, while the origin of the small numbers obtained in the fit to fermion masses is also explained. In our construction, we introduce two clockwork chains consisting of $16+\overline{16}$ vector-like fermions, with all three families coupling to these clockwork chains indistinguishably. The longer of the two chains is responsible for generating small masses for the first generation fermions, while the shorter chain produces the required hierarchy for the second generation. While the clockwork mechanism has been applied separately to address quark flavor puzzle and neutrino masses and mixings, here we attempt a unified description involving both, which occurs naturally in grand unified theories, as in the minimal Yukawa sector of Eq. (\ref{yuk}).

The rest of the paper is organized as follow. In Sec.\ \ref{sec2} we show how to consistently implement clockwork mechanism in $SO(10)$ GUTs with the minimal Yukawa sector. In Sec.\ \ref{sec3}, we build complete models for both the SUSY and non-SUSY $SO(10)$ scenarios. In Sec.\ \ref{sec4} we perform a numerical analysis of the complete fermion spectrum and demonstrate the predictivity of the setup. Finally, we conclude in Sec.\ \ref{sec5}.  Two Appendices contains some technical details.

%%%%%%%%%%%%%%%%%%%%%%%%%%%%%%%%%%%%%%%%%%%%%%%
%%%%%%%%%%%%%%%%%%%%%%%%%%%%%%%%%%%%%%%%%%%%%%%
\section{Clockwork \texorpdfstring{$SO(10)$}{SO(10)}: Setup and formalism}\label{sec2}
In this section, we develop a clockwork extended  $SO(10)$ GUT framework, which is equally applicable to  scenarios with or without SUSY.  In the latter case, a Peccei-Quinn symmetry plays the role that holomorphy of the superpotential plays in the case of SUSY. The Yukawa sector of the minimal renormalizable  $SO(10)$ GUT  \cite{Aulakh:1982sw,Clark:1982ai,Aulakh:2003kg, Bajc:2004xe}  consists of $10_H$ and $\overline{126}_H$ Higgs multiplets that interact  with the three generations of fermions $16_i$  ($i=1,2,3$) leading to the Yukawa interactions written in Eq.\ \eqref{yuk}. To address the origin of hierarchical structure, we introduce a clockwork sector that consists of  two sets of vector-like fermions in the $16+\overline{16}$ representations\footnote{For alternative attempts to explain the flavor puzzle utilizing vector-like fermions see Refs. \cite{Babu:1995hr,Babu:1995fp, Babu:1995uu, Babu:2016cri,Babu:2016aro}.}.
One such clockwork chain consists of $n_1$ vector-like pairs $\chi_a+\overline{\chi}_a$ ($a= 1, ..., n_1$) charged under a  $U(1)_1$ symmetry. The second clockwork chain contains $n_2$  vector-like pairs $\psi_b+\overline{\psi}_b$ ($b= 1, ..., n_2$) charged under a separate $U(1)_2$ symmetry. These abelian symmetries are broken by the VEVs of two separate scalar fields (flavons) $\phi_1$ and $\phi_2$ that are singlets of $SO(10)$ group.  
The $U(1)_1 \times U(1)_2$ charges of all particles involved in the fermion mass generation are listed in Table \ref{tab1}.
Although the $U(1)$ symmetries may be taken to be global, the charge assignments are anomaly-free, and therefore the $U(1)$s can be identified as true gauge symmetries, which is what we shall adopt here.

%\FloatBarrier
\begin{table}[th!]
\centering
\scalemath{0.92}
{
\begin{tabular}{c|c|c|c|c|c|c|c|c|c}
&
$16_i$&
$10_H$&
$\overline{126}_H$&
$\chi_a\; (16)$&$\overline{\chi}_a\; (\overline{16})$&
$\psi_b\; (16)$&$\overline{\psi}_b\; (\overline{16})$& 
$\phi_1\;(1)$&
$\phi_2\; (1)$
\\ [1ex] \hline

$U(1)_1\times U(1)_2$&$(0,0)$&$(0,0)$&$(0,0)$&$(+a,0)$&$(-a,0)$&$(0,+b)$&$(0,-b)$&$(+1,0)$&$(0,+1)$

\end{tabular}}
\caption{Charges of the fields relevant for fermion mass generation under $U(1)_1\times U(1)_2$ symmetry.}\label{tab1}
\end{table}

\begin{figure}[th!]
\centering
\includegraphics[scale=0.32]{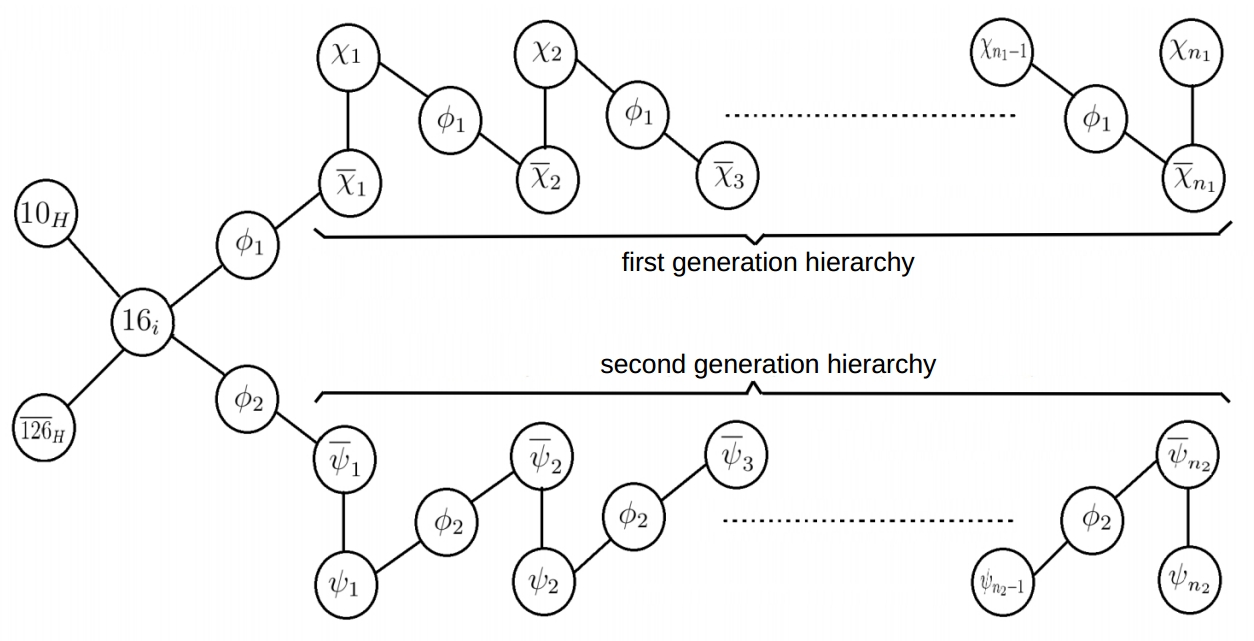}
\caption{ Schematic diagram for the clockwork mechanism in the proposed $SO(10)$ GUT. Solid lines correspond to the interactions present in the theory.  Dotted lines represent the repetition of the clockwork chains. } \label{fig1}
\end{figure}

Then the most general Yakawa interactions consistent with the $U(1)_1 \times U(1)_2$ symmetry are given by
\begin{align}
\mathcal{L}_Y&= 16^T_i\left(Y^{ij}_{10}\,10_H+Y^{ij}_{126}\overline{126}_H\,\right)16_j +\sum_{a=1}^{n_1}  M_a\; \chi_a\overline{\chi}_a + \sum_{b=1}^{n_2} \overline{M}_b\; \psi_b\overline{\psi}_b 
\nonumber \\&
+\sum_{a=2}^{n_1} y_a\;\phi_1\; \chi_{a-1}  \overline{\chi}_a
+\sum_{b=2}^{n_2} \overline{y}_b\;\phi_2\; \psi_{b-1}  \overline{\psi}_b
+\phi_1\; \alpha_i\; 16_i\; \overline{\chi}_1
+\phi_2\; \beta_i\; 16_i\; \overline{\psi}_1\;.
\label{spot}
\end{align}
Note  the interesting fact that all three generations of $16_i$ fermions couple to the clockwork chains indistinguishably. We shall see that a hierarchical structure will arise even in this case. Now, without loss of generality by making rotations in the flavor space one can bring the vectors $\alpha_i$ and $\beta_i$ to the following forms:
\begin{align}
 &\alpha= (\alpha_1,0,0),\;\;\;   \beta= (\beta_1,\beta_2,0). \label{vec}
\end{align}
Furthermore, for the sake of simplicity, we assume universal coupling strength along  each chain that are taken to be real, and define the following quantities 
\begin{align}
&M_a=M_1,\;\;\alpha_1 \langle \phi_1 \rangle = y_a \langle \phi_1 \rangle \equiv -q_1 M_1,\\
&\overline{M}_b=M_2,\;\;\beta_2 \langle \phi_2 \rangle = \overline{y}_b \langle \phi_2 \rangle \equiv -q_2 M_2,
\;\;\beta_1 \langle \phi_2 \rangle \equiv -q_3 M_2. 
\end{align}
Note that with this choice,  $\beta_1$ is the only term that couples the two clockwork chains.We first discuss a scenario where the two clockwork chains are decoupled,  with $\beta_1=0$, and later present the most general analysis with $\beta_1\neq 0$. A schematic diagram to understand our proposed clockwork mechanism in $SO(10)$ GUT is presented in Fig.\ \ref{fig1}.

%%%%%%%%%%%%%%%%%%%%%%%%%%%%%%%%%%%%%%%%%%%%%%%
%%%%%%%%%%%%%%%%%%%%%%%%%%%%%%%%%%%%%%%%%%%%%%%

%%%%%%%%%%%%%%%%%%%%%%%%%%%%%%%%%%%%%%%%%%%%%%%
%%%%%%%%%%%%%%%%%%%%%%%%%%%%%%%%%%%%%%%%%%%%%%%
\subsection{Case with \texorpdfstring{$\beta_1= 0$}{beta1=0}}\label{beta0}
First we analyse the decoupled  scenario of the two different clockwork chains that corresponds to $\beta_1=0$. In this simplified version of the theory, Eq.\ \eqref{spot} takes the following simple form:
\begin{align}
\mathcal{L}_Y=  16^T_i\left(Y^{ij}_{10}\,10_H+Y^{ij}_{126}\overline{126}_H\;\right)16_j &+ M_1 \sum_{a=1}^{n_1} \left( \chi_a\overline{\chi}_a -  q_1\; \chi_{a-1}  \overline{\chi}_a \right) 
%\nonumber \\&
+M_2 \sum_{b=1}^{n_2} \left( \psi_b\overline{\psi}_b -q_2\;  \psi_{b-1}  \overline{\psi}_b \right). \label{spot-ii}
\end{align}
In the above Yukawa interactions, the only two fermions $16_1$ and $16_2$ that directly couple to $10_H$ and $\overline{126}_H$ Higges, we denote them by $\chi_0$ and $\psi_0$, respectively. First consider the chain associated to  $\chi$ fields. The corresponding mass matrix  written in a basis $\overline{\chi} \mathcal{M}_{\chi} \chi$ with $\overline{\chi}= \left( \overline{\chi}_1, ... ,\overline{\chi}_{n_1} \right)$ and $\chi=\left( \chi_0, \chi_1, ... , \chi_{n_1} \right)^T$    has the following form: 
\begin{align}
\mathcal{M}_{\chi}=U \mathcal{M}^{diag}_{\chi}V^T=M_1\begin{pmatrix}
-q_1 & 1 & 0 & \cdots &  & 0 \cr
0 & -q_1 & 1 & \cdots &  & 0 \cr
0 & 0 & -q_1 & \cdots & & 0 \cr
\vdots & \vdots & \vdots & \ddots & &\vdots \cr
% & & & & 1 & 0 \cr
 0 & 0 & 0 &\cdots & -q_1 & 1
\end{pmatrix} _{n_1\times (n_1+1)}\;. \label{toeplitz}
\end{align}
The above matrix  is digonalized by the $U$ and $V$ matrices that are unitary. The eigenvalues of this matrix, including one zero mode and  $n_1$ non-zero states are given by \cite{book}
\begin{align}
m^2_0=0;\;\; m^2_a=M_1^2\left( 1+q_1^2-2q_1\cos\left[ \frac{a \pi}{n_1+1} \right] \right); \;\; a= 1,...,n_1.  \label{cweig}  
\end{align}
We are interested in cases with $q_i\gtrsim \mathcal{O}(1)$; then the mass gap between two consecutive states is of order $\mathcal{O}(M_i)$ as can be seen from Eq.\ \eqref{cweig}. The analysis performed in this section is very general, and thus we need not  specify the mass scale of the vector-like fermions. We will discuss this in more detail in the next section where we present complete models.    The $n_1\times n_1$ unitary matrix $U$ has elements given by \cite{book}
\begin{align}
U_{ja}=-\sqrt{ \frac{2}{n_1+1} } \sin\left[ \frac{(n_1-j+1)\pi}{n_1+1} \right];\;\; j, a= 1,...,n_1.    
\end{align}
Furthermore, the elements of the $(n_1+1)\times (n_1+1)$ unitary matrix $V$ are as follows \cite{book}:
\begin{align}
&V_{j0}=\sqrt{\frac{q_1^2-1}{q_1^2-q_1^{-2n_1}}}\; q_1^{j-n_1}; \;\; j=0,....,n_1,\;a= 1,...,n_1, \label{v1}\\
&V_{ja}=\sqrt{\frac{2}{n_1+1}\frac{M^2}{m^2_a}} \left( q_1\sin\left[ \frac{(n_1-j)a\pi}{n_1+1} \right] - \sin\left[ \frac{(n_1-j+1)a\pi}{n_1+1} \right] \right). \label{v2}
\end{align}
From Eq.\ \eqref{v1}, it is clear that the  massless mode is $\chi'_0\equiv \hat{16}_1$ and that the only field  $\chi_0\equiv 16_1$ that coupled originally to the SM Higgs in Eq.\ \eqref{spot-ii} (contained in the first two terms) are related by 
\begin{align}
%&\chi_0=V_{00}\chi'_0\\
&16_1=V_{00} \hat{16}_1 + \cdots \nonumber
\\ 
&= \hat{16}_1\; \sqrt{\frac{q_1^2-1}{q_1^2-q_1^{-2n_1}}}\;\frac{1}{q_1^{n_1}}\equiv \e \;\hat{16}_1 \label{eq1}.
\end{align}
Here $\cdots$ in the first line represents additional contributions from the heavy fields, which for our purpose are not important, and thus omitted in the next line. 
Since the SM fermions are contained in the zero-modes, the above equation demonstrates that the corresponding Yukawa couplings will have a suppression factor of order  $\e$ associated with the first generation fermions for $q_1>1$. This suppression can even be exponential, provided that  $q_1, n_1 \gg 1$, although to explain flavor hierarchy these numbers need to be only somewhat larger than 1.

By repeating the whole process for the second decoupled chain containing $\psi$ fields, one obtains a similar expression  for the massless mode  $\psi'_0\equiv \hat{16}_2$ and the original field  $\psi_0\equiv 16_2$ 
\begin{align}
%&\psi_0=V'_{00}\psi'_0\\
%&16_2=V'_{00} \hat{16}_2= \hat{16}_2\; \sqrt{\frac{{q'}^2-1}{{q'}^2-{q'}^{-2n'}}}\;\frac{1}{{q'}^{n'}} \equiv \epsilon_2 \;\hat{16}_2. \label{eq2}
&16_2= \hat{16}_2\; \sqrt{\frac{q_2^2-1}{q_2^2-q_2^{-2n_2}}}\;\frac{1}{q_2^{n_2}} \equiv \ee \;\hat{16}_2. \label{eq2}
\end{align}
Hence the Yukawa couplings associated with the second generation fermions receive a suppression of order $\ee \ll 1$ provided that $q_2>1$. These suppression factors are the origin of the fermion mass hierarchies. Assuming both $q_1$ and $q_2$ not very much larger than 1, the length of the clockwork chain associated with $\chi$ fields is required to be longer than the corresponding chain with the $\psi$ fields in order to generate the correct mass hierarchy between the first and the second families. Thus we choose  $n_1>n_2$.

Now, integrating out the heavy fields and making use of Eqs.\ \eqref{eq1} and \eqref{eq2}, we obtain for the light fermion Yukawa couplings as
\begin{align}
\mathcal{L}_Y&=  \hat{16}^T \Lambda^T\; \left(Y_{10}\,10_H+Y_{126}\overline{126}_H\,\right) \;\Lambda\; \hat{16},\;\;~~~~~~~{\rm where}~~~ \Lambda=\begin{pmatrix}
\e&&\\
&\ee&\\
&&1
\end{pmatrix},\label{eqW}
\end{align}
with $\hat{16}=\left( \hat{16}_1\; \hat{16}_2\; \hat{16}_3 \right)^T$ and where the obvious identification $\hat{16}_3= 16_3$ has been  made. This $\Lambda$ matrix entering from the clockwork sector is the origin behind the observed hierarchical pattern of the  fermion masses and mixings. This analysis shows that the Yukawa sector of the theory has the same number of parameters as in the minimal $SO(10)$ model, but with the couplings $Y^{ij}_{10},\, Y^{ij}_{126}$ being of order one. The mass and mixing hierarchies arise from the clockwork chains. The  re-definitions $Y_{10} \rightarrow \Lambda^T Y_{10} \Lambda$ and $Y_{126} \rightarrow \Lambda^T Y_{126} \Lambda$ shows that the parameter count remains the same as in Eq. (\ref{yuk}). For the convenience of the reader, we display the modified form of the Yukawa couplings explicitly:
\begin{align}
&\Lambda^T Y_{k}\Lambda=\begin{pmatrix}
Y^{11}_{k}\;\es&Y^{12}_{k}\;\e\ee&Y^{13}_{k}\;\e&\\ 
Y^{12}_{k}\;\e\ee&Y^{22}_{k}\;\ees&Y^{23}_{k}\;\ee&\\
Y^{13}_{k}\;\e&Y^{23}_{k}\;\ee&Y^{13}_{k}&
\end{pmatrix},\;\;Y^{ij}_{k}\sim \mathcal{O}(1),\;\;k=10,\;126. \label{Ypattern}   
\end{align}
As will be detailed in Sec. \ref{sec4}, the minimal Yukawa couplings of $SO(10)$ that reproduce the correct masses and mixings for both the charged fermions as well as the neutrinos have the unique  hierarchical structure as that of Eq.\ (\ref{Ypattern}), which we will utilize for our numerical study. In Sec. \ref{sec4},  we perform a numerical analysis of the fermion masses and mixings and present our results for the down-quark and up-quark mass matrices  in Eqs.\ \eqref{mdSS} - \eqref{muSS} and in Eqs.\ \eqref{mdNS} - \eqref{muNS}
 for SUSY and non-SUSY scenarios, respectively.

%%%%%%%%%%%%%%%%%%%%%%%%%%%%%%%%%%%%%%%%%%%%%%%
%%%%%%%%%%%%%%%%%%%%%%%%%%%%%%%%%%%%%%%%%%%%%%%
\subsection{Case with \texorpdfstring{$\beta_1\neq  0$}{beta1=/=0}}\label{betano0}
Generically, the term $\beta_1$ in Eq.\ \eqref{vec} is non-zero. In this section, we discuss this general case and consider the following Yukawa interactions:
\begin{align}
\mathcal{L}_Y&=  16^T_i\left(Y^{ij}_{10}\,10_H+Y^{ij}_{126}\overline{126}_H\,\right)16_j + M_1 \sum_{a=1}^{n_1} \left( \chi_a\overline{\chi}_a -  q_1\; \chi_{a-1}  \overline{\chi}_a \right)
\nonumber \\ &
+M_2 \sum_{b=1}^{n_2} \left( \psi_b\overline{\psi}_b -q_2\;  \psi_{b-1}  \overline{\psi}_b \right)  - q_3 M_2 \chi_0 \overline{\psi}_1\;. \label{spot-iii}
\end{align}
Compared to Eq.\ \eqref{spot-ii}, the newly added term is the very last entry in Eq.\ \eqref{spot-iii}. This is the only term that couples the two clockwork chains, due to which, in this section we will follow a different method to compute the effective Yukawa Lagrangian.  As before, we are interested in finding the overlap of the massless modes ($\hat{16}_i$)  with the $0$-th site ($16_i$), which we achieve by integrating out heavy fields one-by-one as discussed below.  Integrating out fermions at the $p$-th site leads to the following relation, which shows the overlap of the $p$-th site with the $(p-1)$-th site:
\begin{align}
&\begin{pmatrix}
\chi_p\\\psi_p
\end{pmatrix}
=L_p
\begin{pmatrix}
\chi_{p-1}\\\psi_{p-1}
\end{pmatrix}.\label{LP}
\end{align}
This way of integrating out the heavy states is convenient for our purpose, however, unlike the previous section,  these transformation matrices $L_p$ are non-unitary. As a result, they modify the kinetic terms, which need to be brought back to the canonical form as done below.   

Since we are interested in the scenario with $n_1>n_2$, $L_p$ appearing in Eq.\ \eqref{LP} has the form 
\begin{align}
&L_p=\begin{cases}
\begin{pmatrix}
q_1&0\\
q_3&q_2
\end{pmatrix};\;\;p=1.
\\
\begin{pmatrix}
q_1&0\\
0&q_2
\end{pmatrix};\;\;p=2,3,...,n_2.
\\
q_1;\;\;p=n_2+1,...,n_1.
\end{cases}
\end{align}
By applying the above definitions, we find the overlap between the $0$-th and the last site as
\begin{align}
&\begin{pmatrix}
\chi_0\\\psi_0\\16_3
\end{pmatrix}
=\left( Q_n ... Q_2 Q_1  \right)^{-1}
\begin{pmatrix}
\chi_{n_1}\\\psi_{n_2}\\16_3
\end{pmatrix}.   \label{over}
\end{align}
Recall that associated with the third generation, there is no clockwork chain, and furthermore $n_1>n_2$. With these conditions, the $3\times 3$ matrices $Q_p$ of Eq. (\ref{over}) are found to be
\begin{align}
&Q_p=\begin{cases}
\begin{pmatrix}
L_p&0\\
0&1
\end{pmatrix};\;\;p=1,2,...,n_2.
\\
\begin{pmatrix}
L_p&0\\
0&\mathcal{I}_{2\times 2}
\end{pmatrix};\;\;p=n_2+1,...,n_1.
\end{cases}
\end{align}
This way of integrating out the clockwork fields modifies the 
kinetic terms that take the following form:
\begin{align}
K\supset \sum_{p=0}^{n_1} \begin{pmatrix} 
\chi_p&\psi_p&16_3
\end{pmatrix}^\dagger
\slashed{D} \begin{pmatrix}
\chi_p\\ \psi_p\\16_3
\end{pmatrix}
= \begin{pmatrix} 
\chi_{n_1}&\psi_{n_2}&16_3
\end{pmatrix}^\dagger
\slashed{D} Z \begin{pmatrix}
\chi_{n_1}\\ \psi_{n_2}\\16_3
\end{pmatrix}\;, \label{kin}
\end{align}
where we have defined 
\begin{align}
&Z=\mathcal{I}_{3\times 3}+\left( P^{-1}_n \right)^{\dagger} \left( P^{-1}_n \right) + ... + \left( P^{-1}_1 P^{-1}_2 ... P^{-1}_n \right)^{\dagger} \left( P^{-1}_1 P^{-1}_2 ... P^{-1}_n \right).
\\
&P^{-1}_p=\begin{cases}
\begin{pmatrix}
L_p^{-1}&0\\
0&0
\end{pmatrix};\;\;p=1,2,...,n_2.
\\
\begin{pmatrix}
L_p^{-1}&0\\
0&0_{2\times 2}
\end{pmatrix};\;\;p=n_2+1,...,n_1.
\end{cases}    
\end{align}
Then canonical normalization of the kinetic terms given in  Eq.\ \eqref{kin} along with Eq.\ \eqref{over} provides the desired relation: 
\begin{align}
\begin{pmatrix}
\chi_0\\\psi_0\\16_3
\end{pmatrix}= \left( Q_n ... Q_2 Q_1  \right)^{-1} Z^{-1/2}  \begin{pmatrix}
\hat{16}_1\\\hat{16}_2\\\hat{16}_3
\end{pmatrix} 
=\Lambda  \begin{pmatrix}
\hat{16}_1\\\hat{16}_2\\\hat{16}_3
\end{pmatrix}. 
\end{align}
More explicitly, the suppression factors that originate from the clockwork sector are embedded in the matrix $\Lambda$ given by
\begin{align}
\Lambda&= \left( Q_n ... Q_2 Q_1  \right)^{-1} Z^{-1/2}
\\&
= \begin{pmatrix}
q_1^{-n_1}&0&0\\
-\frac{q_3}{q_2}q_1^{-n_1}&{q_2}^{-n_2}&0\\
0&0&1
\end{pmatrix} \begin{pmatrix}
\frac{q_1^2-q_1^{-2n_1}}{q_1^2-1}&0&0\\
-\frac{q_3}{q_2}q_1^{-2n_1}& \frac{{q_2}^2-{q_2}^{-2n_2}}{{q_2}^2-1}&0\\
0&0&1
\end{pmatrix}^{-1/2} \label{repro}
\\&
= \begin{pmatrix}
q_1^{-n_1}&0&0\\
-\frac{q_3}{q_2}q_1^{-n_1}&{q_2}^{-n_2}&0\\
0&0&1
\end{pmatrix}  K_1  \begin{pmatrix}
\lambda_1^{-1/2}&0&0\\
0&\lambda_2^{-1/2}&0\\
0&0&1
\end{pmatrix}  K_2^T  \label{lam2b}
\\&
\equiv \begin{pmatrix}
\epsilon_{11}&\epsilon_{12}&0\\
\epsilon_{21}&\epsilon_{22}&0\\
0&0&1
\end{pmatrix}. \label{lam2}
\end{align}
The matrices $K_{1,2}$  as well as the eigenvalues $\lambda_{1,2}$ are defined in Appendix \ref{A}.  As can be seen from Eq.\ \eqref{repro}, with $q_3=0$ (corresponding to two decoupled chains), one reproduces Eqs.\ \eqref{eq1} and \eqref{eq2}.

%%%%%%%%%%%%%%%%%%%%%%%%%%%%%%%%%%%%%%%%%%%%%%%
%%%%%%%%%%%%%%%%%%%%%%%%%%%%%%%%%%%%%%%%%%%%%%%
\section{Model implementation}\label{sec3}
%%%%%%%%%%%%%%%%%%%%%%%%%%%%%%%%%%%%%%%%%%%%%%%
%%%%%%%%%%%%%%%%%%%%%%%%%%%%%%%%%%%%%%%%%%%%%%%
In Sec.\ \ref{sec2}, we have discussed the clockwork implementation of the minimal Yukawa sector of $SO(10)$ without being specific to the theory being supersymmetric or not.  In this section, we provide the necessary details to implement the mechanism in complete models with and without SUSY.

%%%%%%%%%%%%%%%%%%%%%%%%%%%%%%%%%%%%%%%%%%%%%%%
%%%%%%%%%%%%%%%%%%%%%%%%%%%%%%%%%%%%%%%%%%%%%%%
\subsection{SUSY \texorpdfstring{$SO(10)$}{SO(10)} model}\label{susy}

In minimal SUSY $SO(10)$ GUT,  in addition to $10_H$ and $\overline{126}_H$, a  $210_H$ Higgs representation is employed to consistently  break the GUT symmetry in the SUSY limit. Proton decay constraints require the GUT symmetry breaking scale to be around $M_{GUT}\sim 10^{16}$ GeV, which is also the scale where the gauge couplings unify in the  minimal supersymmetric standard model (MSSM). On the other hand, to generate viable light neutrino masses via type-I seesaw mechanism,  the right-handed neutrinos must have masses which are a few orders smaller than the GUT scale, requiring  $v_R\sim 10^{12}-10^{13}$ GeV ($v_R$ is the VEV of the SM singlet component of $126_H$). 
In this minimal setup, such a low value of $v_R$ would lead to certain colored states from the $126_H$ acquiring intermediate scale masses, thus spoiling  perturbative gauge coupling unification \cite{Bertolini:2006pe,Aulakh:2005bd, Bajc:2005qe}.    

A simple choice to solve this issues is to introduce a $54_H$  Higgs multiplet  \cite{Babu:2018tfi}, which can break $SO(10)$ down to $SU(4)_c \times SU(2)_L \times SU(2)_R$ symmetry. It also supplies GUT scale masses to the would-be light colored states. Since $54_H$ has no couplings to fermion bilinears, the minimal Yukawa sector of Eq. (\ref{yuk}) remains intact, which is our foucs here.

The $210_H$ can have renormalizable couplings with the fermions belonging to the  clockwork sector of the form  $\chi_a\overline{\chi}_a \;210_H$ (and similarly $\psi_b\overline{\psi}_b \;210_H$). The presence of these terms would introduce some modifications to the analysis performed in the previous section. This can be easily avoided by imposing  a $\mathcal{Z}_4$ discrete symmetry. The full charge assignment that can do the job is presented in Table.\ \ref{tabz}.   Note that with this charge assignment the bare mass terms of the vector-like fermions  would break the $\mathcal{Z}_4$ to a $\mathcal{Z}_2$.  Alternatively, the VEV of a flavon field carrying -2 units of $Z_4$ charge break  $\mathcal{Z}_4$ spontaneously to $\mathcal{Z}_2$.

%\FloatBarrier
\begin{table}[th!]
\centering
\scalemath{0.88}
 {
\begin{tabular}{c|c|c|c|c|c|c|c|c|c|c}
&
$16_i$&
$10_H$&
$\overline{126}_H$&$210_H$&
$\chi_a\; (16)$&
$\overline{\chi}_a\; (\overline{16})$&
$\psi_b\; (16)$& 
$\overline{\psi}_a\; (\overline{16})$&
$\phi_1\; (1)$& 
$\phi_2\; (1)$
\\ [1ex] \hline

$\mathcal{Z}_4$&+1&-2&-2&0&+1&+1&+1&+1&-2&-2
\\ \hline

$U(1)_1\times U(1)_2$&$(0, 0)$&$(0, 0)$&$(0, 0)$&$(0, 0)$&$(+a, 0)$&$(-a, 0)$&$(0, +b)$&$(0, -b)$&$(+1, 0)$&$(0, +1)$

\end{tabular}}
\caption{Charges of the particle of SUSY $SO(10)$ model with the imposition of a $\mathcal{Z}_4$ symmetry.  Here $\mathcal{Z}_4$ charges are defined such that $\omega^{q_4}=1$ for $q_4=4$, where $\omega=e^{i2\pi/4}$. The model also contains $\overline{\phi}_1$ and $\overline{\phi}_2$ fields with charges opposite to those of $\phi_{1,2}$.   }\label{tabz}
\end{table}

In this set-up with the $\mathcal{Z}_4$ symmetry, for our numerical study presented later in the text, we shall consider a case with $n_1=4$ and $n_2=2$. The mass scale of these vector-like fermions will be taken to be above the GUT scale. Consequently, the successful perturbative  gauge coupling unification of MSSM would remain intact\footnote{As is well known, the minimal SUSY $SO(10)$ model has large beta function coefficients of order $\mathcal{O}(100)$ for gauge coupling evolution above the GUT scale, and the theory begins to become non-perturbative  at mass scales $\mu$ few times larger than the GUT scale, but below the Planck scale $M_{pl}$.  Ways to deal with the non-perturbative nature of the theory in this momentum scale have been discussed in Ref. \cite{Aulakh:2003kg, Bajc:2004xe, Aulakh:2002ph}. In our theory, the introduction of six vector-like pairs of $16+\bar{16}$ adds $+24$ to this already large beta function coefficient (If an additional $210$ of Higgs is used symmetry breaking, the $b$-factor changes from $+109$ to $+133$).  Hence, keeping the clockwork fields above the GUT scale affects the gauge coupling evolution above the GUT scale only minimally.} \cite{Dimopoulos:1981yj, Ibanez:1981yh, Einhorn:1981sx, Marciano:1981un}.

%%%%%%%%%%%%%%%%%%%%%%%%%%%%%%%%%%%%%%%%%%%%%%%
%%%%%%%%%%%%%%%%%%%%%%%%%%%%%%%%%%%%%%%%%%%%%%%
\subsubsection{Symmetry breaking of \texorpdfstring{$U(1)_i$}{U(1)}}
In this subsection, we discuss the symmetry breaking of  $U(1)_i$ under which only the clockwork fields carry non-zero charges. We follow the method developed in Ref. \cite{Fayet:1974pd,Fayet:1975yi,Fayet:1978ig} for achieving $U(1)$ gauge symmetry breaking in the supersymmetric limit, taking advantage of the Fayet-Iliopoulos term allowed for abelian symmetries \cite{Fayet:1974jb}. 
The breaking of $U(1)_i$ should be achieved by flavon superfields $\phi_i+\overline{\phi}_i$ that carry  $Q(\phi_i)=-q_i$ and $Q(\overline{\phi}_i)=+q_i$ charges under the respective  $U(1)_i$. Then one immediately realizes that in the superpotential given in Eq.\ \eqref{spot}, a term of the form $\overline{\phi}_1 \overline{\chi}_{a-1}\chi_a$ (and $\overline{\phi}_2 \overline{\psi}_{b-1}\psi_b$) must be added. Such a term would spoil the successful implementation of the clockwork mechanism, and must be suppressed. This can be achieved if  the VEV of $\overline{\phi}_i$ is significantly smaller than the VEV of $\phi_i$. Here we show that these fields can have completely different VEVs $\langle \phi_i \rangle \neq \langle \overline{\phi}_i \rangle$ \cite{Tavartkiladze:2011ex}. Now to fix all the VEVs and lift the flat directions, we introduce one more scalar $S_i$ which is neutral under $U(1)_i$. Then the relevant  superpotential can be written as \cite{Fayet:1974pd}:
\begin{align}
W_{U(1)_i}= \lambda_i S_i(\phi_i\overline{\phi}_i-\mu_i^2)  \;,  \label{Fterm} 
\end{align}
where, $\lambda_i$ is a dimensionless parameters. In addition, the superpotential also contains terms that are quadratic and cubic in $S_i$. Since the symmetry under consideration is abelian, in general  a Fayet-Iliopoulos \cite{Fayet:1974jb} term, which is both SUSY and gauge invariant is allowed in the Lagrangian  that has the form $\xi_i \int d^4\theta V_{U(1)_i}$, where $\xi_i$ a parameter that has dimension of mass$^2$. The associated $D$-term, upon integrating out the auxiliary component, has the form
\begin{align}
 D_{U(1)_i}=\xi_i -q_i|\phi_i|^2 +q_i|\overline{\phi}_i|^2\;. \label{Dterm}  
\end{align}
In the unbroken SUSY limit, both the $F$-terms and the $D$-terms must vanish, which from Eqs.\ \eqref{Fterm} and \eqref{Dterm} can be written as
\begin{align}
\lambda_i (\phi_i\overline{\phi}_i-\mu_i^2)=0,\;\; \lambda_i S_i\overline{\phi}_i=0,\;\; \lambda_i S_i\phi_i=0,\;\;  \xi_i -q_i|\phi_i|^2 +q_i|\overline{\phi}_i|^2=0 \;.
\end{align}
These relations have the following solution:
\begin{align}
S_i=0,\;\;    \phi_i\overline{\phi}_i= \mu_i^2, \;\; \frac{\xi_i}{q_i}= |\phi_i|^2 -|\overline{\phi}_i|^2\;,
\end{align}
from which the VEVs of the flavon fields can be fixed as
\begin{align}
|\phi_i|^2=\frac{1}{\sqrt{2}} \left[ \frac{\xi_i}{q_i}+\left(  \frac{\xi_i^2}{q_i^2} + 4 |\mu_i|^2 \right)^{1/2}  \right], \;\;
|\overline{\phi}_i|^2= \frac{\sqrt{2}|\mu_i|^2}{\frac{\xi_i}{q_i}+\left(  \frac{\xi_i^2}{q_i^2} + 4 |\mu_i|^2 \right)^{1/2}}\;.
\end{align}
Then our desired VEV structure can be archived in the following limit\footnote{There is another limit where  $|\phi_i|\ll |\overline{\phi}_i|$ with   $\frac{\xi_i}{q_i}<0$ and $|\mu_i|^2\ll - \frac{\xi_i}{q_i}$, which we are not interested in.}
\begin{align}
&\frac{\xi_i}{q_i}>0,\;\;\; \frac{\xi_i}{q_i}\gg |\mu_i|^2,
\\
&|\phi_i|=\sqrt{\frac{\xi_i}{q_i}},\;\;\; |\overline{\phi}_i|= |\mu_i|^2 \sqrt{\frac{q_i}{\xi_i}}\;;\;\;\;\Rightarrow \;\; |\phi_i|\gg |\overline{\phi}_i|\;.
\end{align}
This justifies the omission of terms containing $\overline{\phi}_i$ superfields in the superpotential Eq.\ \eqref{spot}.  Thus, the fermion mass fits arising from the minimal Yukawa sector of Eq.\ \eqref{spot} is realized within the model, with the clockwork mechanism explaining the hierarchical patterns.

%%%%%%%%%%%%%%%%%%%%%%%%%%%%%%%%%%%%%%%%%%%%%%%
%%%%%%%%%%%%%%%%%%%%%%%%%%%%%%%%%%%%%%%%%%%%%%%
\subsection{Non-SUSY \texorpdfstring{$SO(10)$}{SO(10)} model}\label{nonsusy}

In the non-SUSY $SO(10)$ GUT, $10_H$ Higgs can be taken to be either real or complex.  However, a real $10_H$ with $\overline{126}_H$ Higgs alone does not lead to a realistic fermion mass spectrum \cite{Bajc:2005zf,Babu:2016bmy}.  If a complex $10_H$ is employed, an additional Yukawa coupling matrix will result from the $10_H^*$. One interesting possibility is to augment $SO(10)$ with a global $U(1)_{PQ}$  PQ  symmetry  as suggested in Ref. \cite{Babu:1992ia} \footnote{For earlier works on the implementation of PQ symmetry in $SO(10)$ GUT, see for example Refs. \cite{Davidson:1983fe, Davidson:1983fy}.}. This would require complexification of the $10_H$, but the $10_H^*$ will not couple to fermions owing to the PQ charge. Introduction of the PQ symmetry is highly motivated, as it solves the strong CP problem, and also provides a dark matter candidate in the form of axion.  There exists two known classes of consistent ``invisible'' axion models: the KSVZ model  \cite{Kim:1979if, Shifman:1979if} and the DFSZ model \cite{Dine:1981rt,Zhitnitsky:1980tq}. In this work, we adopt the KSVZ axion model that suits well with the clockwork setup.   We assume the existence of an $SO(10)$ singlet scalar $\phi_0$ that carries nonzero charge under $U(1)_{PQ}$, whose VEV  breaks the PQ symmetry spontaneously.

Now, to reproduce the analysis performed in Sec.\ \ref{sec2}, and to achieve the same suppression for fermion masses from the clockwork sectors, in this non-SUSY framework the vector-like fermions must carry charges under the PQ symmetry. Our chosen charge assignments of  fields under $U(1)_{PQ}$ and $U(1)_1\times U(1)_2$ are presented in Table.\ \ref{tab2}. A charge assignment of this type forbids the unwanted terms involving $\phi^*_{1,2}$ in the Yukawa Lagrangian which helps regain the true clockwork nature. With these charges, the complete Yukawa Lagrangian including the clockwork chains practically  has the same form as that of Eq.\ \eqref{spot}. Consequently, the analysis performed in Secs. \ref{beta0} and  \ref{betano0} remain valid. The most general Yakawa Lagrangian consistent with all symmetries has the  form:
\begin{align}
\mathcal{L}_Y&= 16^T_i\left(Y^{ij}_{10}\,10_H+Y^{ij}_{126}\overline{126}_H\,\right)16_j +\sum_{a=1}^{n_1}  \lambda_a\;\phi_0 \chi_a\overline{\chi}_a + \sum_{b=1}^{n_2} \overline{\lambda}_b\;\phi_0 \psi_b \overline{\psi}_b 
\nonumber \\&
+\sum_{a=2}^{n_1} y_a\;\phi_1\; \chi_{a-1}  \overline{\chi}_a
+\sum_{b=2}^{n_2} \overline{y}_b\;\phi_2\; \psi_{b-1}  \overline{\psi}_b
+\phi_1\; \alpha_i\; 16_i\; \overline{\chi}_1
+\phi_2\; \beta_i\; 16_i\; \overline{\psi}_1.
\label{lag}
\end{align}
Note that, as opposed to Eq.\ \eqref{spot}, in the above Lagrangian, there is no bare mass  for the vector-like fermions. The clockwork fields get their masses only after the PQ symmetry breaks. Following the same notation as in Eq.\ \eqref{spot}, we identify $\lambda_a\langle \phi_0\rangle \equiv M_a$ and $\overline{\lambda}_b\langle \phi_0\rangle \equiv \overline{M}_b$. We emphasise that our particular chosen charge assignments automatically  forbids couplings involving $\phi^*_{1,2}$ to preserve the clockwork nature of the Lagrangian.

%\FloatBarrier
\begin{table}[th!]
\centering
\scalemath{0.85}
 {
\begin{tabular}{c|c|c|c|c|c|c|c|c|c|c}
&
$16_i$&
$10_H$&
$\overline{126}_H$&
$\chi_a\; (16)$&
$\overline{\chi}_a\; (\overline{16})$&
$\psi_b\; (16)$& 
$\overline{\psi}_a\; (\overline{16})$&
$\phi_1\; (1)$& 
$\phi_2\; (1)$&
$\phi_0\; (1)$
\\ [1ex] \hline

$U(1)_{PQ}$&+1&-2&-2&+1&+1&+1&+1&-2&-2&-2
\\ \hline

$U(1)_1\times U(1)_2$&$(0, 0)$&$(0, 0)$&$(0, 0)$&$(+a, 0)$&$(-a, 0)$&$(0, +b)$&$(0, -b)$&$(+1, 0)$&$(0, +1)$&$(0, 0)$

\end{tabular}}
\caption{$U(1)$ charges of the particles of non-SUSY $SO(10)$ model.}\label{tab2}
\end{table}

Since the vector-like fermions in the clockwork sector receive their masses only after PQ symmetry breaking, these fields have masses below the GUT scale and  contribute to the beta function coefficients of renormalization group equations (RGEs) for the gauge couplings in the momentum range $f_{PQ} \leq \mu \leq M_{GUT}$, whre $f_{PQ}$ is the PQ symmetry breaking scale. This however, does not change the unification of the    gauge couplings  of the minimal $SO(10)$ GUT with an intermediate scale \cite{Rizzo:1981dm, Rizzo:1981jr, Caswell:1982fx, Chang:1983fu, Gipson:1984aj, Chang:1984qr, Deshpande:1992au, Deshpande:1992em, Bertolini:2009qj, Bertolini:2009es, Bertolini:2010ng, Babu:2015bna, Graf:2016znk, Babu:2016bmy, Chakrabortty:2019fov, Meloni:2019jcf}, but only change the value of the unified gauge coupling at the GUT scale, as the clockwork chains form complete $SO(10)$ multiplets.  To show the consistency of our model, in the following  we consider gauge coupling unification for two different cases with six pairs of vector-like in the  $16+\overline{16}$ representation having masses of order the PQ scale.  For simplicity of our analysis, we take them to be degenerate and fix their common masses at the PQ scale.  In the first scenario, we assume that a $54_H$ Higgs breaks $SO(10)$ down to the Pati-Salam (PS) symmetry $SU(2)_L \times SU(2)_R \times SU(4)_c$ at the GUT scale. In this case, we evolve the one-loop  SM RGEs for the gauge couplings with well known SM beta function coefficients $b_i=\{41/10,-19/6,7\}$ \cite{Jones:1981we} from low scale $M_Z$ to PQ scale $M_{PQ}$ that we fix to be $10^{12}$ GeV. At this scale contributions from six pairs of vector-like fermions are added, which corresponds to $b_i=\{201/10,77/6,9\}$. With these new beta function coefficients, running is done up to the Pati-Salam  scale $M_{PS}$, where proper matching conditions corresponding to PS symmetry with $D$-parity  are imposed. In this procedure we have inputted the low scale  (experimental central) values of the couplings to be $\alpha^{-1}_1(M_Z)= 59.02$, $\alpha^{-1}_1(M_Z)= 29.57$, and $\alpha^{-1}_1(M_Z)= 8.44$ \cite{Antusch:2013jca}, which gives us the PS scale to be $M_{PS}=4.9\times 10^{13}$ GeV.   Now, for the consistency of symmetry breaking as well as for generating realistic fermion spectrum, the case under investigation requires the entire $\overline{126}_H$ multiplet and a complex $(2,2,1)\subset 10_H$ to have masses  at the PS scale. From this scale  we  evolve the new PS gauge couplings with beta function coefficients  $b_i=\{74/3,17\}$ (here $i=1,2$ correspond to $SU(2)_{L}$, and $SU(4)_C$ respectively) up to the GUT scale $M_{GUT}$, where unification is demanded.   

%%%%%%%%%%%%%%%%%%%%%%%%%%%%%%%%%%%%%%%%%%%%%%%
\begin{figure}[t!]
\centering
\includegraphics[scale=1]{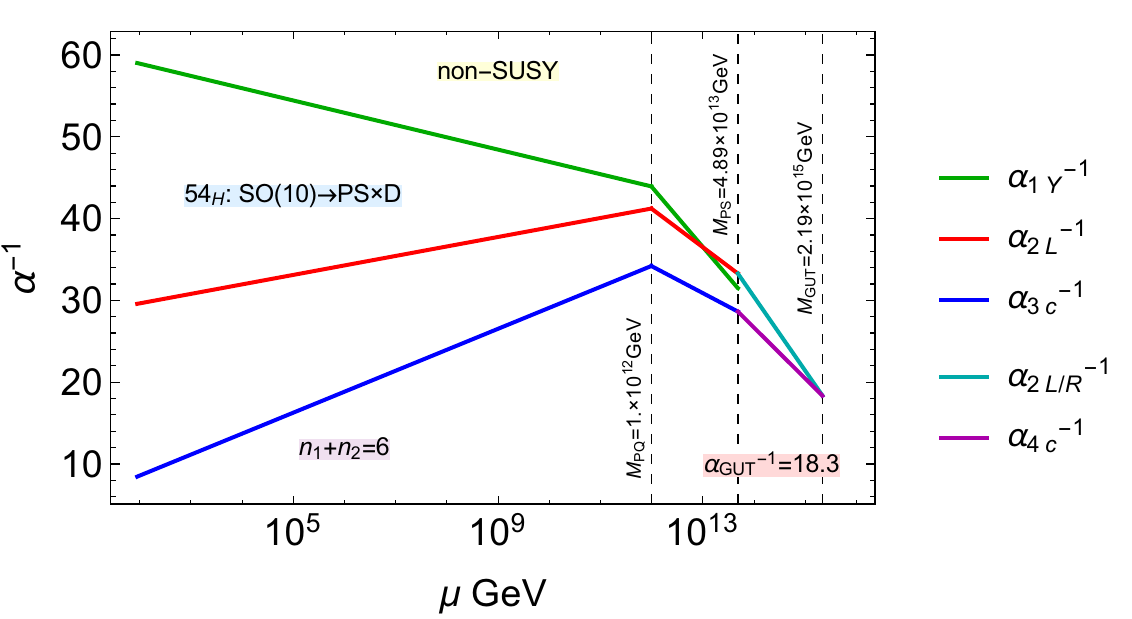}
\includegraphics[scale=1]{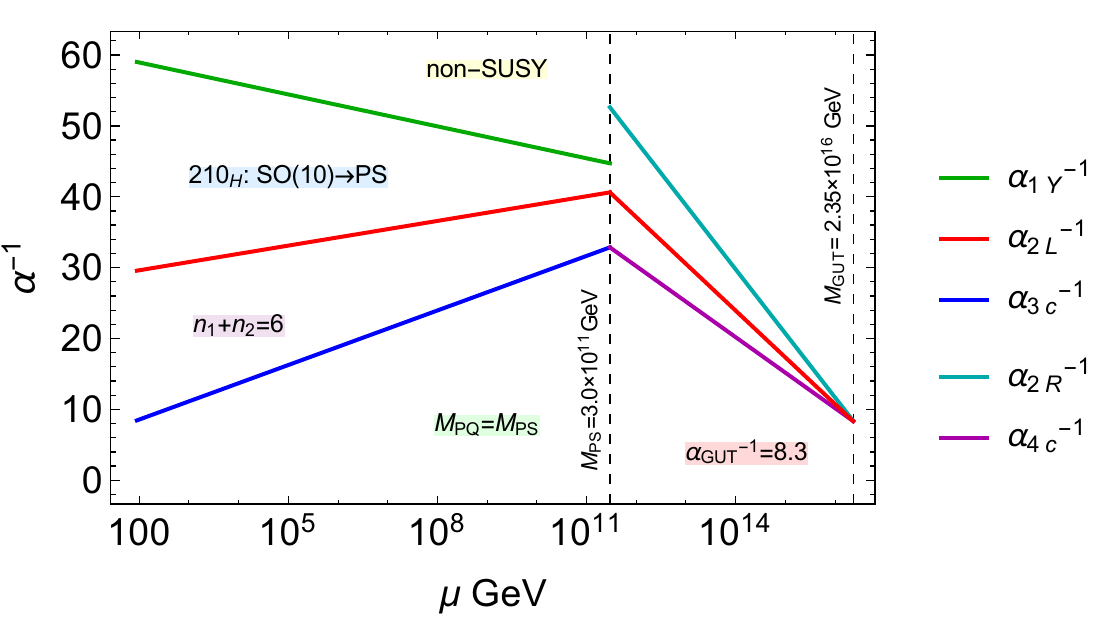}
\caption{One-loop gauge coupling unification for non-SUSY $SO(10)$ model for two different scenarios: (i) symmetry breaking with $54_H$ Higgs (upper plot), (ii) symmetry breaking with $210_H$ without $D$-parity (lower plot). In both cases six pairs of vector-like fermions are kept at the PQ scale. } \label{fig2}
\end{figure}
%%%%%%%%%%%%%%%%%%%%%%%%%%%%%%%%%%%%%%%%%%%%%%%

In the second case, we achieve the GUT symmetry breaking via $210_H$ Higgs, and assume the absence of  $D$-parity \cite{Chang:1985zq} at the PS scale. In this case we also take the PQ and the PS breaking scale to be the same,  $M_{PQ}=M_{PS}$. Moreover, above  the intermediate scale gauge  beta functions receive  contributions from the clockwork sector as before,  as well as contributions from  $(2,2,15), (1,3,\overline{10}) \subset \overline{126}_H$, and a complex $(2,2,1)\subset 10_H$ of Higgs bosons.  The beta function coefficients are  $b_i=\{18,74/3,41/3\}$ (here $i=1,2,3$ correspond to $SU(2)_L$, $SU(2)_R$, and $SU(4)_C$ respectively). 

Perturbative gauge coupling unification can be obtained in each of the aforementioned scenarios, and these results are presented in Fig.\ \ref{fig2}. We have used one-loop RGE and ignored high scale threshold effects in generating these figures.
From Fig.\ \ref{fig2}, one clearly sees the advantage of employing a $210_H$ Higgs. First of all, it is possible in this case to identify the PS scale with the PQ scale.  Furthermore, unification occurs at around $2.35 \times 10^{16}$ GeV, which implies that proton lifetime arising from gauge boson mediated processes is sufficiently long.  The larger unification scale is correlated with the smaller intermediate PS scale, which is realized since the $210_H$ breaks $SO(10)$ down to the PS symmetry without $D$ parity.   When a $54_H$ is used to break $SO(10)$ symmetry instead of the $210_H$,  one sees that the unification scale is relatively low, about $2\times 10^{15}$ GeV. This is correlated with a larger intermediate PS scale, which is a consequence of an unbroken $D$-parity.  It is this symmetry that requires the $SU(2)_L$ partner of the $(10,1,3)$ Higgs multiplet to have mass at the PS scale, thus affecting the gauge coupling evolution more drastically.  A null observation of proton decay requires the GUT scale to be $M_{GUT}\geq 5\times 10^{15}$ GeV. It has been shown that including high scale threshold corrections this model can indeed be consistent with proton lifetime limits \cite{Babu:2015bna}.  It should also be noted that a $45_H$ can be used to 
break the GUT symmetry, but the viability of this scenario relies  on quantum corrections in the Higgs potential
\cite{Graf:2016znk}.  We note that the analysis performed in the Yakawa sector remains valid regardless of the choice of the Higgs field that breaks the GUT symmetry.

In the case of non-SUSY $SO(10)$ embedded with clockwork chain, the gauge couplings remain perturbative all the way to the Planck scale.  The beta-function coefficient $b$, where $dg/dt = b g^3/(16 \pi^2)$, changes from $-34/3$ to $+14/3$ with the addition of six pairs of $16+\bar{16}$ of the clockwork sector as an example.  (This choice is not unique, however.)  This leaves the gauge coupling perturbative up to the Planck scale. Here we assumed that an additional $54$ Higgs is involved in $SO(10)$ symmetry breaking.

%%%%%%%%%%%%%%%%%%%%%%%%%%%%%%%%%%%%%%%%%%%%%%%
\begin{figure}[t!]
\centering
\includegraphics[scale=1]{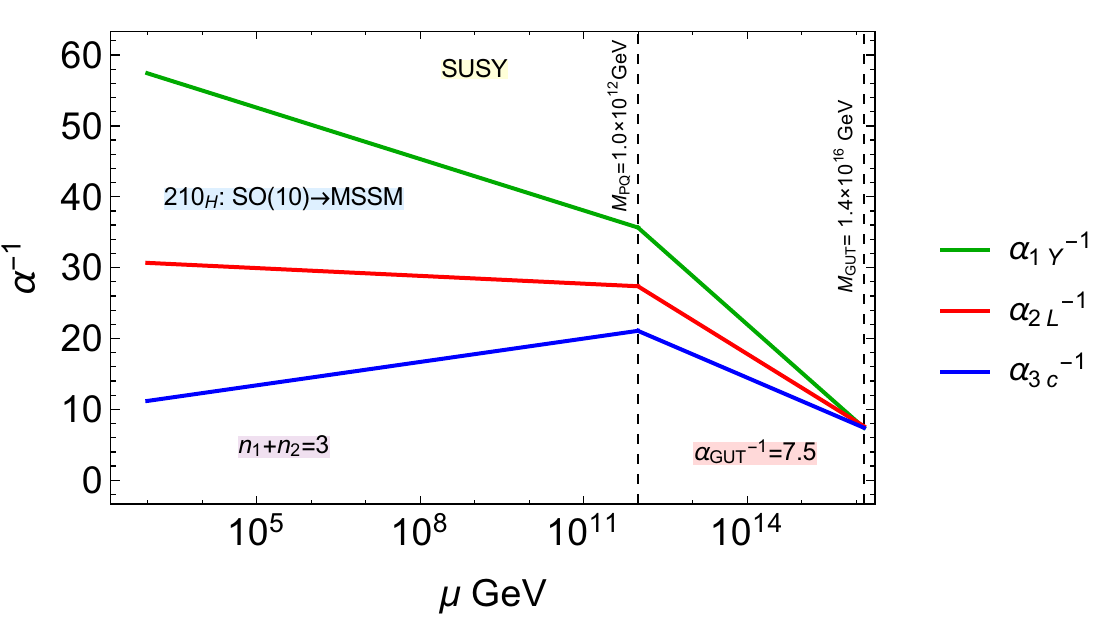}
\caption{Gauge coupling unification for SUSY $SO(10)$ model with three sets of  vector-like $16+\overline{16}$  fermions  added at the PQ scale.    For this illustration, we set $M_{SUSY}= 1$ TeV and used the one-loop RGE evolution of gauge couplings.} \label{fig3}
\end{figure}
%%%%%%%%%%%%%%%%%%%%%%%%%%%%%%%%%%%%%%%%%%%%%%%

%%%%%%%%%%%%%%%%%%%%%%%%%%%%%%%%%%%%%%%%%%%%%%%
%%%%%%%%%%%%%%%%%%%%%%%%%%%%%%%%%%%%%%%%%%%%%%%
\subsection{SUSY \texorpdfstring{$SO(10)$}{SO(10)} with PQ symmetry}\label{susyPQ}

The Peccei-Quinn symmetry can be implemented in the SUSY $SO(10)$ framework along the line discussed in the previous subsection.\footnote{In this work, we do not discuss the details of the PQ symmetry breaking and refer the reader  to Ref. \cite{Babu:2018qca} for successful implementation of $U(1)_{PQ}$ in the context of minimal  SUSY $SO(10)$ GUT.} With the PQ symmetry, the unwanted couplings of the $210_H$ field with the vector-like fermions will be forbidden, and there would be no need for a $Z_4$ symmetry adopted in Sec. \ref{susy}. This however, modifies the successful gauge coupling unification of the MSSM.  The reason is that, in this set-up, the vector-like fermions acquire their masses only after the PQ symmetry is broken. Hence, above the  PQ scale, the beta function coefficients for the gauge coupling evolution receive additional contributions. Consequently, perturbative unification of gauge couplings even up to the GUT scale becomes challenging. We have checked that unlike the non-SUSY case, adding six pairs  of vector-like fermions at the PQ scale certainly does not work for the SUSY scenario. In Fig.\ \ref{fig3}, we demonstrate a viable perturbative one-loop gauge coupling unification scenario, with three vector-like fermion pairs  $16+\overline{16}$ having masses at the PQ scale. In this analysis, we  evolve the MSSM  RGEs from the SUSY scale ($M_{SUSY}=$ 1 TeV) up to the PQ scale with the MSSM beta function coefficients $b_i=\{33/5,1,-3\}$. The input values at the TeV scale for the gauge couplings are taken to be  $\alpha^{-1}_1(M_{SUSY})= 57.43$, $\alpha^{-1}_1(M_{SUSY})= 30.67$, and $\alpha^{-1}_1(M_{SUSY})= 11.19$ \cite{Antusch:2013jca}. Then at  the PQ scale, we add contributions from three pairs of vector-like fermions that modifies the new beta function coefficients to be  $b_i=\{93/5,13,9\}$ and further run the RGEs up to the GUT scale. Whereas  Fig.\ \ref{fig3} shows the consistency of keeping three vector-like pairs at the intermediate scale,  adding any more pairs would bring the theory to a nonperturbative regime before reaching the GUT scale. Since the implementation of clockwork mechanism typically requires larger number of vector-like states, this scenario with the added PQ symmetry is not a preferred option, since it makes clockwork not very efficient. Furthermore, due to the presence of the plethora
of fields lurking around the GUT scale, large threshold corrections are expected to play vital role in the running of the gauge couplings (see for example Ref. \cite{Aulakh:2013lxa}), which we have not taken into consideration.  However, even this scenario can explain partially the hierarchies in the fermion and mixing angles.

%%%%%%%%%%%%%%%%%%%%%%%%%%%%%%%%%%%%%%%%%%
%%%%%%%%%%%%%%%%%%%%%%%%%%%%%%%%%%%%%%%%%%
\section{Fit to the fermion spectrum}\label{sec4} 

%%%%%%%%%%%%%%%%%%%%%%%%%%%%%%%%%%%%%%%%%%
%%%%%%%%%%%%%%%%%%%%%%%%%%%%%%%%%%%%%%%%%%
\subsection{SUSY case}
To fit the fermion masses and mixings, we perform a $\chi^2$ analysis, for which we closely follow the procedure discussed in detail in Refs. \cite{Babu:2018tfi, Babu:2018qca}. From Eq.\ \eqref{eqW}, first we obtain the fermion mass matrices which have the following form:
\begin{align}
&M_d=\Lambda^T (H+F)\Lambda \;, \label{mat1}\\
&M_u=r \Lambda^T (H+s F)\Lambda \;,\\
&M_e=\Lambda^T (H-3 F)\Lambda \;,\\
&M^{\nu}_D=r \Lambda^T (H- 3 s F)\Lambda \;,\\
&M_R=r_R \Lambda^T F\Lambda \;,\\
&M_N=-{M^{\nu}_D}^T M_R^{-1} M^{\nu}_D \;, \label{mat2}
\end{align}
where we have defined
\begin{align}
&r= \frac{v^{10}_u}{v^{10}_d}, \;\;\; s=\frac{1}{r} \frac{v^{126}_u}{v^{126}_d}, \;\;\; r_R= \frac{v_R}{v^{126}_d},  \label{rs} \\
&H= v^{10}_{d} Y_{10}, \;\;\;
 F= v^{126}_{d}Y_{126}\;.
\end{align}
Here $v_u^{10}$ is the VEV of the up-type Higgs doublet from $10_H$, etc.
As in Ref. \cite{Babu:2018tfi}, these mass matrices are written in  $f^cM_f f$ basis.  Note that in Eq.\ \eqref{mat2},  the type-II contributions to neutrino mass is omitted, since the weak triplets have masses of order the GUT scale, hence the corresponding type-II contributions are negligible.

In our numerical analysis, we will take these ratios $r$ and $s$ given in Eq. \eqref{rs} to be free parameters. Note however that, in a theory where the Higgs sector is completely specified, these ratios are related to the parameters that appear in the Higgs potential. Since the focus of the present work is on the {\it minimal Yukawa sector}, rather than a minimal and complete symmetry breaking sector, we do not make such a connection here. Detailed studies along this line have been made in Refs. \cite{Aulakh:2006hs,Aulakh:2008sn,Aulakh:2013lxa,Babu:2018tfi,Babu:2018qca}.

It should be pointed out that due to the presence of the right-handed neutrinos that have masses a few orders less than the GUT scale, the running of the  RGEs from the $M_Z$ scale to the $M_{GUT}$ get modified. We properly include these corrections to the Yukawa couplings due to the intermediate scale threshold; for details of this implementation we refer the reader to Ref. \cite{Babu:2018tfi}. Furthermore, we take the GUT scale values of the charged fermion masses and CKM mixing parameters from Ref. \cite{Babu:2018qca}. In this procedure, the Yukawa couplings, the CKM parameters, and the $d=5$ effective operator for neutrino masses and mixings are run from the $M_Z$ scale to the SUSY scale, which is chosen to be 1 TeV.\footnote{When $54_H$ Higgs is added alongside  $210_H$ Higgs as in Ref. \cite{Babu:2018tfi}, consistency of proton lifetime requires a mini-split SUSY spectrum with the sfermions having masses of order 100 TeV, accompanied by TeV scale gauginos and Higgsinos. In this case, RGEs running will be somewhat different, which corresponds to slightly different input values of the observables at the GUT scale.} Above this scale, the full MSSM RGEs are used up to the GUT scale. For neutrinos, on the contrary, the $d=5$ effective operator running is carried out up to the intermediate scale. For neutrinos, we have taken the low energy values from the recent global fit performed in Ref. \cite{deSalas:2020pgw}. These input values are collected in the second column in Table.\ \ref{result}.

We remind the reader that the original Yukawa couplings $Y_{10}$ and $Y_{126}$ are allowed to have entries that are $\sim \mathcal{O}(1)$, whereas the hierarchies among different generations are generated via the suppression factors $\e$ and $\ee$ originating from the clockwork sector. It is crucial to understand that compared to the minimal $SO(10)$ Yukawa sector, our setup does not introduce any new parameters into the theory. In our fit, we fix $\tan\beta=v_u/v_d= 10$, where  $v_u$ and $v_d$ are the VEVs of the MSSM fields $H^{MSSM}_u$ and $H^{MSSM}_d$.  With these,  we find an excellent fit that corresponds to $\chi^2=7.98$; the best fit values and the associated pulls are presented in the third and the fourth columns of Table.\ \ref{result}. The full set of best fit parameter values can be found in Appendix \ref{C}. From this fit, we chose the following suppression factors
\begin{align}
\Lambda = \begin{pmatrix}
\e&0&0\\
0&\ee&0\\
0&0&1
\end{pmatrix} =  \begin{pmatrix}
0.06117&0&0\\
0&0.32983&0\\
0&0&1
\end{pmatrix}.  
\end{align}
Here $\e\simeq \lambda^2$ and $\ee\simeq \lambda$, with $\lambda=\sin \theta_c$, where $\theta_c$ is the Cabibbo angle.  A choice of these values for $\e$ and $\ee$ correspond to $q_1=1.93445,\; q_2=1.55098$ for $n_1=4,\;n_2=2$.  The down-quark and up-quark mass matrices (after taking into account the intermediate scale threshold corrections) corresponding to the best fit are then given below, which explicitly demonstrates how clockwork is responsible for generating the required hierarchies in the fermion masses and mixings.   All other mass matrices can be readily obtained from these two matrices (or from the parameter set given in Appendix \ref{C}). 
\begin{align}
&M_d=0.80166e^{2.18422 i}\begin{pmatrix}
1.27815\;\es\;e^{-3.0004 i}&1.26174\;\e\ee\;e^{-2.71759 i}&1.11777\;\e\;e^{1.63891 i}\\
1.26174\;\e\ee\;e^{-2.71759 i}&1.61178\;\ees\;e^{-2.31606 i}&1.20224\;\ee\;e^{2.02843 i}\\
1.11777\;\e\;e^{1.63891 i}&1.20224\;\ee\;e^{2.02843 i}&1
\end{pmatrix}   \;\text{GeV}, \label{mdSS} 
\\
&M_u=79.1467e^{2.27241 i}\begin{pmatrix}
1.09514\;\es\;e^{3.10025 i}&1.17556\;\e\ee\;e^{-2.80578 i}&1.04143\;\e\;e^{1.55072 i}\\
1.17556\;\e\ee\;e^{-2.80578 i}&1.24674\;\ees\;e^{-2.42765 i}&1.12012\;\ee\;e^{1.94025 i}\\
1.04143\;\e\;e^{1.55072 i}&1.12012\;\ee\;e^{1.94025 i}&1
\end{pmatrix}\;\text{GeV}. \label{muSS}
\end{align}
It can be seen from the above matrices that all the Yukawa couplings are very close to 1. Even thought in writing Eqs.\ \eqref{mdSS} and \eqref{muSS} we have used the clockwork chain lengths to be $n_1=4$ and $n_2=2$, however, from the best fit parameters  presented in Appendix \ref{C} various different chain lengths can be considered without needing to modify the fit at all. By demanding order unity Yukawa couplings, this corresponds to longer chain length for $q_i$ very close to 1, and shorter chain length when $q_i$ starts to become larger than 1.   Finally, the analysis presented here can be trivially extended  for the case $\beta_1\neq 1$ discussed in Sec.\ \ref{betano0}. In the case of SUSY $SO(10)$ with a Peccei-Quinn symmetry, concentrating on the gauge coupling unification scenario presented in Fig.\ \ref{fig3}, where we have assumed $n_1+n_2=3$,  setting $n_1=2$ and $n_2=1$ returns $q_1=3.9779$ and $q_2=2.8622$, which are not too far from unity.

Note that the parameters of this class of models cannot be chosen randomly. There is a GUT sum rule involving the masses and mixings in the model. This should not be considered as a fine-tuning of parameters, rather it is a result of a GUT symmetry. This has been noted in the Ref. \cite{Babu:1992ia}, and the sum rule has been explicitly worked out in Ref.  \cite{Lavoura:1993vz}  for the case of real parameters. Indeed, in the case of all parameters being real, one can write $M_\ell = a M_u + b M_d$, with  the mass matrices being symmetric.  In a basis where $M_u$ is diagonal, $M_d = V.M_d^{\rm diagonal}.V^T$, where $V$ is the CKM matrix.  Thus all elements of the charged lepton mass matrix $M_\ell$ is determined in terms of the quark masses, CKM mixing angles and the two free parameters $(a,\,b)$.  This leads to one sum rule involving the masses of the charged fermions. Even with the
inclusion of phases the GUT mass sum rule is present in the model, the corresponding relation becomes more involved.  However, for any given set of masses and mixings the magnitudes as well as the phases are constrained, such that this mass sum rule is satisfied. This constraint is implemented numerically in our fits, and as a result, the order one complex parameters in Eqs.\ \eqref{mdSS} and \eqref{muSS} cannot be arbitrarily chosen (the same argument goes for the fit associated to Eqs.\ \eqref{mdNS} and \eqref{muNS}).  This mass sum rule is a prediction of $SO(10)$ GUT, which prevails in our clockwork extension.

Before closing this section we compare the fit we obtain here with the fit presented in Ref. \cite{Babu:2018qca}. In the present work the total $\chi^2$ obtained is about 8, whereas, it is about 6 in  Ref. \cite{Babu:2018qca}. The reason for obtaining a slightly higher $\chi^2$ is mainly due to the fact that in this work in the fitting procedure, we have included the Dirac phase in the lepton sector as well, which was left out in the $\chi^2$-minimization in Ref. \cite{Babu:2018qca}. Additionally, in the prsent work, we have taken the  recent global fit values of the neutrinos, which have somewhat smaller experimental uncertainties compared to the previously used values in Ref. \cite{Babu:2018qca}.

%\newpage
\FloatBarrier
\begin{table}[t!]
\centering
\footnotesize
\resizebox{1\textwidth}{!}{
\begin{tabular}{|c|c|c|c|c|c|c|}
\hline
\textbf{Observables} & \multicolumn{3}{c|}{\bfseries SUSY} 
 & \multicolumn{3}{c|}{\bfseries non-SUSY} \\ 

\cline{2-7}
(masses in GeV) &Input&Best Fit&Pull &Input&Best Fit&Pull  \\
\hline

\rowcolor{red!10}$m_u/10^{-3}$&0.502$\pm$0.155&0.515&0.08&0.442$\pm$0.149&0.462&0.13\\  
\rowcolor{red!10}$m_c$&$0.245\pm$0.007&0.246&0.14&0.238$\pm$0.007&0.239&0.18\\ 
\rowcolor{red!10}$m_t$&90.28$\pm$0.89&90.26&-0.02&74.51$\pm$0.65&74.47&-0.05\\

\rowcolor{yellow!10}$m_b/10^{-3}$&0.839$\pm$0.17&0.400&-2.61&1.14$\pm$0.22&0.542&-2.62\\ 
\rowcolor{yellow!10}$m_s/10^{-3}$&$16.62\pm$0.90&16.53&-0.09&21.58$\pm$1.14&22.57&0.86\\ 
\rowcolor{yellow!10}$m_b$&$0.938\pm$0.009&0.933&-0.55&0.994$\pm$0.009&0.995&0.19\\

\rowcolor{cyan!10}$m_e/10^{-3}$&0.3440$\pm$0.0034&0.344&0.08&0.4707$\pm$0.0047&0.470&-0.03\\ 
\rowcolor{cyan!10}$m_\mu/10^{-3}$&72.625$\pm$0.726&72.58&-0.05&99.365$\pm$0.993&99.12&-0.24\\ 
\rowcolor{cyan!10}$m_\tau$&1.2403$\pm$0.0124&1.247&0.57&1.6892$\pm$0.0168&1.688&-0.05\\

\rowcolor{orange!10}$|V_{us}|/10^{-2}$&$22.54\pm$0.07&22.54&0.02&22.54$\pm$0.06&22.54&0.06\\ 
\rowcolor{orange!10}$|V_{cb}|/10^{-2}$&3.93$\pm$0.06&3.908&-0.42&4.856$\pm$0.06&4.863&0.13\\ 
\rowcolor{orange!10}$|V_{ub}|/10^{-2}$&0.341$\pm$0.012&0.341&0.003&0.420$\pm$0.013&0.421&0.10\\ 
\rowcolor{orange!10}$\delta_{CKM}^{\circ}$&69.21$\pm$3.09&69.32&0.03&69.15$\pm$3.09&70.24&0.35\\

\rowcolor{green!10}$\Delta m^2_{21}/10^{-5} (eV^2)$&8.982$\pm$0.25&8.972&-0.04&12.65$\pm$0.35&12.65&-0.01\\ 
\rowcolor{green!10}$\Delta m^2_{31}/10^{-3} (eV^2)$&3.05$\pm$0.04&3.056&0.02&4.307$\pm$0.059&4.307&0.006\\

\rowcolor{blue!10}$\sin^2 \theta_{12}$&0.318$\pm$0.016&0.314&-0.19&0.318$\pm$0.016&0.316&-0.07\\ 
\rowcolor{blue!10}$\sin^2 \theta_{23}$&0.563$\pm$0.019&0.563&0.031&0.563$\pm$0.019&0.563&0.01\\ 
\rowcolor{blue!10}$\sin^2 \theta_{13}$&0.0221$\pm$0.0006&0.0221&-0.003&0.0221$\pm$0.0006&0.0220&-0.16\\  
\rowcolor{blue!10}$\delta_{CP}^{\circ}$&224.1$\pm$33.3&240.1&0.48&224.1$\pm$33.3&225.1&0.03\\

\hline
\rowcolor{gray!10}$\chi^2$&-&-&7.98&-&-&7.96 \\
\hline

\end{tabular}
}
\caption{Inputs and the corresponding best fit values of the observables along with their pulls at the GUT scale $\mu=2\times 10^{16}$ GeV for both SUSY and non-SUSY cases are summarized here. In both these cases, type-I seesaw dominance is assumed,  for details see text.    }\label{result}
\end{table}

%%%%%%%%%%%%%%%%%%%%%%%%%%%%%%%%%%%%%%%%%%
%%%%%%%%%%%%%%%%%%%%%%%%%%%%%%%%%%%%%%%%%%
\subsection{Non-SUSY case}

To get the GUT scale values of the fermion masses and mixings for the non-SUSY scenario, we closely follow the procedure discussed in Ref. \cite{Babu:2016bmy}. In this method, the low scale values are  evolved up to the GUT scale using SM RGEs. However, this one-step RGE running receives corrections due to the intermediate scale right-handed neutrinos. In our numerical fit, we take into account these modification of the Yukawa couplings following the method detailed in Ref. \cite{Babu:2016bmy}, where  a basis of $fM_{ij}f^c$ is used, and we stay with such a basis. Then, for the non-SUSY case, the mass matrices Eqs.\ \eqref{mat1} - \eqref{mat2}  derived in the previous section are still applicable with the only exception that $M_\nu^D$ should be transposed in Eq.\ \eqref{mat2}.   As before, we focus on the type-I dominance scenario for the neutrino masses. It is to be pointed out that type-II seesaw for non-SUSY case fails to provide a realistic fit \cite{Joshipura:2011nn}. The GUT scale inputs for charged fermion masses and mixings are obtained from  Ref. \cite{Babu:2016bmy}, whereas for neutrinos, we have collected the  recent low scale values from Ref. \cite{deSalas:2020pgw}, and evolved the $d=5$ effective operator up to the right-handed neutrino mass scale. These input parameters are summarized in the fifth column of Table.\ \ref{result}. A good fit to these data is obtained from our numerical procedure, and the the best fit corresponds to $\chi^2=7.96$.  The best fit values for  physical quantities  and their pulls are presented in the sixth and the seventh columns of Table.\ \ref{result}.  The theory parameters of this best fit are summarized in Appendix \ref{D}. As before, no new parameters enter in this fit process compared to the minimal $SO(10)$ Yukawa sector. Hence, our fit is applicable for cases with or without clockwork extension (for both SUSY and non-SUSY models). Following our previous analysis, we fix the clockwork chain lengths to be $n_1=4,\;n_2=2$, then the corresponding chosen suppression factors are
\begin{align}
\Lambda = \begin{pmatrix}
\e&0&0\\
0&\ee&0\\
0&0&1
\end{pmatrix} =  \begin{pmatrix}
0.03783&0&0\\
0&0.23973&0\\
0&0&1
\end{pmatrix}.  
\end{align}
Like in the SUSY case, here we also get suppression factors of same order: $\e\simeq \lambda^2$ and $\ee\simeq \lambda$ that amount to $q_1=2.20312,\; q_2=1.89219$. The associated down-quark and up-quark mass matrices (after taking into account the intermediate scale threshold corrections) are found to be 
\begin{align}
&M_d=0.94968e^{-1.68947 i}\begin{pmatrix}
1.45089\;\es\;e^{2.18915 i}&0.97791\;\e\ee\;e^{2.37807 i}&1.10034\;\e\;e^{1.04364 i}\\
0.97791\;\e\ee\;e^{2.37807 i}&1.22764\;\ees\;e^{2.18995 i}&0.89334\;\ee\;e^{1.09213 i}\\
1.10034\;\e\;e^{1.04364 i}&0.89334\;\ee\;e^{1.09213 i}&1
\end{pmatrix}\;\text{GeV},  \label{mdNS}
\\
&M_u=71.505e^{1.23138 i}\begin{pmatrix}
1.11401\;\es\;e^{2.51117 i}&0.91309\;\e\ee\;e^{2.5988 i}&1.0274\;\e\;e^{1.26437 i}\\
0.91309\;\e\ee\;e^{2.5988 i}&0.73257\;\ees\;e^{2.69088 i}&0.83413\;\ee\;e^{1.31287 i}\\
1.0274\;\e\;e^{1.26437 i}&0.83413\;\ee\;e^{1.31287 i}&1
\end{pmatrix}\;\text{GeV}.\label{muNS}
\end{align}
All the conclusion we have reached for the SUSY case are also applicable in this non-SUSY scenario. 

From the results presented in the previous subsection as well as in this subsection, it is clear that our simple clockwork extension to the minimal Yukawa sector of $SO(10)$ with or without SUSY naturally explains the hierarchies in the fermion spectrum.

%%%%%%%%%%%%%%%%%%%%%%%%%%%%%%%%%%%%%%%%%%
%%%%%%%%%%%%%%%%%%%%%%%%%%%%%%%%%%%%%%%%%%
\section{Conclusion}\label{sec5}
In this work, we have presented a minimal and highly predictive mechanism to address the \textit{flavor puzzle}.  In particular, our proposed framework is based on the minimal Yukawa sector of $SO(10)$ GUT with or without SUSY, extended with two clockwork chains. Each of these chains consists of a set of $16+\overline{16}$ vector-like fermions that couples indistinguishably with different fermion generations.  Whereas $SO(10)$ symmetry correlates different fermion sectors, clockwork sector supplies  proper  suppression factors to incorporate the required hierarchies. The proposed setup to explain the origin of flavor hierarchies is simple in its construction and is also   renormalizable. All  Yukawa couplings of these theories are of order unity which is shown to provide a consistent fit to the fermion masses and mixings. Detailed numerical analysis has been carried out, and the results summarized in Table \ref{result} to demonstrate the robustness  of the theory.

%%%%%%%%%%%%%%%%%%%%%%%%%%%%%%%%%%%%%%%%%%
%%%%%%%%%%%%%%%%%%%%%%%%%%%%%%%%%%%%%%%%%%
\section*{Acknowledgments}
The work of KSB was supported in part by US Department of Energy Grant Number DE-SC 0016013.

\section*{Appendix}
\appendix
%%%%%%%%%%%%%%%%%%%%%%%%%%%%%%%%%%%%%%%%%%%%%%%
\section{Expressions for \texorpdfstring{$\lambda_i$}{lambda} and \texorpdfstring{$K_{1,2}$}{K1,K2}}\label{A}
In this Appendix we present the exact analytical forms for $\lambda_i$ and $K_{1,2}$ matrices as defined in Eq.\ \eqref{lam2b}:
\begin{align}
&\begin{pmatrix}
\frac{q_1^2-q_1^{-2n_1}}{q_1^2-1}&0&0\\
-\frac{q_3}{q_2}q_1^{-2n_1}& \frac{{q_2}^2-{q_2}^{-2n_2}}{{q_2}^2-1}&0\\
0&0&1
\end{pmatrix}
\equiv 
\begin{pmatrix}
a&0&0\\b&c&0\\0&0&1
\end{pmatrix}=K_1 \begin{pmatrix}
\lambda_1&0&0\\0&\lambda_2&0\\0&0&1
\end{pmatrix} K^T_2.
\\&
\lambda_{1,2}=\left[    
\frac{1}{2} \left(   
a^2+b^2+c^2\mp \sqrt{a^2+b^2+c^2-4 a^2c^2}
\right)
\right]^{1/2}.
\\&
K_1=\left(
\begin{array}{ccc}
 -\frac{-a^2+b^2+c^2+\sqrt{a^4+2 \left(b^2-c^2\right)
   a^2+\left(b^2+c^2\right)^2}}{2 a b
   \sqrt{\frac{\left(-a^2+b^2+c^2+\sqrt{a^4+2 \left(b^2-c^2\right)
   a^2+\left(b^2+c^2\right)^2}\right)^2}{4 a^2 b^2}+1}} &
   -\frac{-a^2+b^2+c^2-\sqrt{a^4+2 \left(b^2-c^2\right)
   a^2+\left(b^2+c^2\right)^2}}{2 a b
   \sqrt{\frac{\left(a^2-b^2-c^2+\sqrt{a^4+2 \left(b^2-c^2\right)
   a^2+\left(b^2+c^2\right)^2}\right)^2}{4 a^2 b^2}+1}}&0 \\
 \frac{1}{\sqrt{\frac{\left(-a^2+b^2+c^2+\sqrt{a^4+2 \left(b^2-c^2\right)
   a^2+\left(b^2+c^2\right)^2}\right)^2}{4 a^2 b^2}+1}} &
   \frac{1}{\sqrt{\frac{\left(a^2-b^2-c^2+\sqrt{a^4+2 \left(b^2-c^2\right)
   a^2+\left(b^2+c^2\right)^2}\right)^2}{4 a^2 b^2}+1}} &0 \\ 0&0&1
\end{array}
\right).
\\&
K_2=\left(
\begin{array}{ccc}
 -\frac{-a^2-b^2+c^2+\sqrt{a^4+2 \left(b^2-c^2\right)
   a^2+\left(b^2+c^2\right)^2}}{2 b c
   \sqrt{\frac{\left(a^2+b^2-c^2-\sqrt{a^4+2 \left(b^2-c^2\right)
   a^2+\left(b^2+c^2\right)^2}\right)^2}{4 b^2 c^2}+1}} &
   \frac{1}{\sqrt{\frac{\left(a^2+b^2-c^2-\sqrt{a^4+2 \left(b^2-c^2\right)
   a^2+\left(b^2+c^2\right)^2}\right)^2}{4 b^2 c^2}+1}} &0\\
 -\frac{-a^2-b^2+c^2-\sqrt{a^4+2 \left(b^2-c^2\right)
   a^2+\left(b^2+c^2\right)^2}}{2 b c
   \sqrt{\frac{\left(a^2+b^2-c^2+\sqrt{a^4+2 \left(b^2-c^2\right)
   a^2+\left(b^2+c^2\right)^2}\right)^2}{4 b^2 c^2}+1}} &
   \frac{1}{\sqrt{\frac{\left(a^2+b^2-c^2+\sqrt{a^4+2 \left(b^2-c^2\right)
   a^2+\left(b^2+c^2\right)^2}\right)^2}{4 b^2 c^2}+1}}& 0 \\ 0&0&1
\end{array}
\right).
\end{align}

%%%%%%%%%%%%%%%%%%%%%%%%%%%%%%%%%%%%%%%%%%%%%%%
\section{Best fit parameters}\label{B}
As discussed in the main text, the Yukawa sector is effectively identical to the minimal $SO(10)$ model. It is because the clockwork sector does not introduce any new parameters, rather it  accounts for the hierarchical factors. Hence, the fit we perform is identical to the minimal Yukawa sector of $SO(10)$ model. Also our fit can be used for arbitrary lengths of the clockwork chains. Due to these attractive features, in the following, for the convenience of the readers, we present our best fit parameters in the form that is readily used for general purpose. Following Refs. \cite{Babu:2016bmy,Babu:2018tfi} we present the best fit parameters in a basis where the $\Lambda^T Y_{126} \Lambda$ is diagonal and real. We have used these best fit parameters to reconstruct the down-quark and up-quark mass matrices, which are presented in Eqs.\ \eqref{mdSS} -  \eqref{muSS} and in Eqs.\ \eqref{mdNS} -  \eqref{muNS} for SUSY and non-SUSY models, respectively.  As can be seen from these mass matrices, all the Yukawa couplings are of the same order and in fact very close to unity, which is our desired result. Note however that besides the Yukawa couplings, a fit to the fermion spectrum contains two VEV ratios $s$ and $r$ as defined in Eq.\ \eqref{rs}. For the former, our fit prefers a value of $s\simeq \lambda$, and for the latter  $r\simeq m_t/m_b$ is required. These VEV ratios do not have any direct connection to the Yukawa couplings and do not necessarily have to be of order unity.  Within our framework, their values are predicted directly from a fit to the data.

%%%%%%%%%%%%%%%%%%%%%%%%%%%%%%%%%%%%%%%%%%%%%%%
\subsection{SUSY \texorpdfstring{$SO(10)$}{SO(10)}}\label{C}
\begin{align}
&r=93.9719,\;
s=2.96269\times 10^{-1} + 1.27201\times 10^{-2} i,\; r_R=8.73689\times 10^{12}\;,\label{S-1}
\\
&\Lambda^T F \Lambda =10^{-1}\left(
\begin{array}{ccc}
 6.59098\times 10^{-3} & 0. & 0. \\
 0. & 3.41720\times 10^{-1} & 0. \\
 0. & 0. & 1.33390 \\
\end{array}
\right)\;\text{GeV},  
\\
&\Lambda^T H \Lambda =10^{-3}\left(
\begin{array}{ccc}
 1.96740\, -2.79338 i & 17.5736\, -10.3763 i &
   -42.5697-34.5327 i \\
 17.5736\, -10.3763 i & 105.17\, -18.4787 i &
   -152.329-279.013 i \\
 -42.5697-34.5327 i & -152.329-279.013 i & -594.884+655.504
   i \\
\end{array}
\right)\;\text{GeV}. \label{S-2}
\end{align}

%%%%%%%%%%%%%%%%%%%%%%%%%%%%%%%%%%%%%%%%%%%%%%%
\subsection{Non-SUSY \texorpdfstring{$SO(10)$}{SO(10)}}\label{D}
\begin{align}
&r=70.3027,\;
s=2.57526\times 10^{-1} + 5.27538\times 10^{-2} i,\; r_R=4.57993\times 10^{12}\;,\label{NS-1}
\\
&\Lambda^T F \Lambda =10^{-1}\left(
\begin{array}{ccc}
 5.29819\times 10^{-3} & 0 & 0 \\
 0 & 3.82033 \times 10^{-1} & 0 \\
 0 & 0 & 3.04637 \\
\end{array}
\right)   \;\text{GeV}, 
\\
&\Lambda^T H \Lambda =10^{-3}\left(
\begin{array}{ccc}
 1.20097\, -0.94480 i & 6.50337\, -5.35207 i & 31.5693\,
   +23.7924 i \\
 6.50337\, -5.35207 i & 20.5847\, -32.1521 i & 168.169\, +114.396
   i \\
 31.5693\, +23.7924 i & 168.169\, +114.396 i & -417.079+943.005 i
   \\
\end{array}
\right) \;\text{GeV}. \label{NS-2}
\end{align}

%%%%%%%%%%%%%%%%%%%%%%%%%%%
\bibliographystyle{style}
\bibliography{ref}
\end{document}